\pdfoutput=1
\documentclass{article}
\PassOptionsToPackage{numbers,sort&compress}{natbib}
\usepackage{lmodern}
\usepackage[preprint]{neurips_2026}
\usepackage[utf8]{inputenc}
\usepackage[T1]{fontenc}
\usepackage{amsmath,amssymb,graphicx,booktabs,array,tabularx,xcolor}
\definecolor{questioncolumn}{RGB}{237,243,248}
\newsavebox{\relatedworktable}
\usepackage[colorlinks=true,allcolors=blue!45!black]{hyperref}
\usepackage{url}
\renewcommand{\arraystretch}{1.12}
\newcolumntype{Y}{>{\raggedright\arraybackslash}X}
\newcolumntype{P}[1]{>{\raggedright\arraybackslash}p{#1}}
\newcommand{\model}[1]{\texttt{#1}}
\newenvironment{introlist}{\begin{list}{\textbullet}{%
  \setlength{\leftmargin}{1.1em}\setlength{\labelwidth}{.7em}\setlength{\labelsep}{.4em}%
  \setlength{\topsep}{0pt}\setlength{\partopsep}{0pt}%
  \setlength{\itemsep}{2pt plus 1pt minus .5pt}\setlength{\parsep}{0pt}}}{\end{list}}

\newcommand{\CoreLower}{0.659}
\newcommand{\CoreUpper}{0.909}

\newcommand{\ExternalScore}{0.984}

\newcommand{\ClaudeRFQCoverageSix}{81.2\%}
\newcommand{\ClaudeRFQCoverageEight}{98.6\%}

\newcommand{\PairedCoreDelta}{0.056}
\newcommand{\PairedAstraCore}{[0.655, 0.912]}
\newcommand{\PairedClaudeCore}{[0.711, 0.968]}
\newcommand{\PairedCoreDifference}{[-0.201, 0.313]}
\newcommand{\PairedAstraExternal}{0.946}
\newcommand{\AstraFirstTurnMinutes}{15.8}
\newcommand{\AstraCompletedTurnMinutes}{10.2}
\newcommand{\ClaudeFirstTurnMinutes}{41.4}
\newcommand{\ClaudeCompletedTurnMinutes}{21.2}

\newcommand{\AstraCoreFlags}{13/32}
\newcommand{\ClaudeCoreFlags}{5/28}
\newcommand{\AstraPairedCoreFlags}{11/28}
\newcommand{\ClaudePairedCoreFlags}{5/28}

\newcommand{\RFQComparisonRows}{%
Codex / Sol & 20 & 63.0 & 80.8 & 11/20 & 3/115 \\
Codex / Astra & 20 & 75.5 & 97.9 & 17/20 & 2/137 \\
Claude Code / Opus (amended) & 20 & 81.2 & 98.6 & 18/20 & 0/136 \\
Claude / original & 17 & 81.2 & 99.2 & 16/17 & 0/116 \\
}

\newcommand{\RFQPairedCaseRows}{%
001 & 71.4 & 100.0 & 100.0 & N/Y/Y \\
002 & 85.7 & 100.0 & 100.0 & Y/Y/Y \\
003 & 71.4 & 100.0 & 100.0 & N/Y/Y \\
004$^{*}$ & 57.1 & 100.0 & 100.0 & N/Y/Y \\
005 & 85.7 & 100.0 & 100.0 & N/Y/Y \\
006 & 85.7 & 85.7 & 100.0 & Y/N/Y \\
007 & 100.0 & 100.0 & 100.0 & Y/Y/Y \\
008 & 71.4 & 85.7 & 85.7 & N/N/N \\
009$^{*}$ & 85.7 & 100.0 & 100.0 & Y/Y/Y \\
010 & 85.7 & 100.0 & 100.0 & Y/Y/Y \\
011 & 100.0 & 85.7 & 100.0 & Y/N/Y \\
012 & 57.1 & 100.0 & 100.0 & N/Y/Y \\
013 & 100.0 & 100.0 & 100.0 & Y/Y/Y \\
014 & 100.0 & 100.0 & 100.0 & Y/Y/Y \\
015 & 85.7 & 100.0 & 100.0 & Y/Y/Y \\
016 & 16.7 & 100.0 & 100.0 & N/Y/Y \\
017 & 85.7 & 100.0 & 100.0 & N/Y/Y \\
018 & 71.4 & 100.0 & 100.0 & N/Y/Y \\
019$^{*}$ & 100.0 & 100.0 & 85.7 & Y/Y/N \\
020 & 100.0 & 100.0 & 100.0 & Y/Y/Y \\
}

\newcommand{\BriefPairedCaseRows}{%
L3-BID-002 & R & [0.900, 1.000] & [0.900, 1.000] & Complete \\
L3-CAM-002 & R & [0.900, 1.000] & Unscored & Timeout \\
L3-CFR-002 & R & [0.500, 0.600] & [0.600, 0.700] & Complete \\
L3-CFRY-001 & R & [0.900, 1.000] & [0.900, 1.000] & Complete \\
L3-ECO-002 & R & [0.900, 1.000] & [0.900, 1.000] & Complete \\
L3-MFA-001 & R & [0.900, 1.000] & Unscored & Timeout \\
L3-PSF-002 & R & [0.900, 1.000] & [0.900, 1.000] & Complete \\
L3-RDT-001 & R & 0.970 & 1.000 & Complete \\
L3-RDT-002 & R & 0.980 & Unscored & Timeout \\
L3-RDT-003 & R & 0.970 & 1.000 & Complete \\
L3-RDT-004 & R & 1.000 & 1.000 & Complete \\
L3-RDT-005 & R & 0.980 & 1.000 & Complete \\
L3-RDT-006 & R & 1.000 & Unscored & Timeout \\
L3-RDT-010 & R & 1.000 & Unscored & Timeout \\
L3-RDT-011 & R & 1.000 & 1.000 & Complete \\
L3-RTO-002 & R & [0.900, 1.000] & [0.900, 1.000] & Complete \\
L4-BID-001 & P & [0.460, 0.760] & [0.700, 1.000] & Complete \\
L4-BID-001 & R & [0.460, 0.760] & Unscored & Timeout \\
L4-BID-001 & Z & [0.400, 0.700] & [0.550, 0.850] & Complete \\
L4-BID-003 & R & [0.700, 1.000] & [0.700, 1.000] & Complete \\
L4-CAM-001 & P & [0.700, 1.000] & [0.550, 0.850] & Complete \\
L4-CAM-001 & R & [0.700, 1.000] & [0.700, 1.000] & Complete \\
L4-CAM-001 & Z & [0.550, 0.850] & [0.700, 1.000] & Complete \\
L4-CAM-003 & R & [0.700, 1.000] & [0.700, 1.000] & Complete \\
L4-CFR-001 & R & [0.700, 1.000] & [0.700, 1.000] & Complete \\
L4-CFR-003 & R & [0.700, 1.000] & [0.700, 1.000] & Complete \\
L4-CFRY-002 & R & [0.430, 0.730] & [0.700, 1.000] & Complete \\
L4-CFRY-003 & R & [0.550, 0.850] & [0.700, 1.000] & Complete \\
L4-ECO-001 & P & [0.700, 1.000] & [0.700, 1.000] & Complete \\
L4-ECO-001 & R & [0.700, 1.000] & [0.610, 0.910] & Complete \\
L4-ECO-001 & Z & [0.700, 1.000] & [0.700, 1.000] & Complete \\
L4-ECO-003 & R & [0.700, 1.000] & [0.700, 1.000] & Complete \\
L4-MFA-002 & R & [0.520, 0.820] & Unscored & Timeout \\
L4-MFA-003 & R & [0.460, 0.760] & [0.490, 0.790] & Complete \\
L4-PSF-001 & R & [0.550, 0.850] & [0.700, 1.000] & Complete \\
L4-PSF-003 & R & [0.700, 1.000] & [0.700, 1.000] & Complete \\
L4-RDT-007 & R & 0.700 & 1.000 & Complete \\
L4-RDT-008 & R & 1.000 & 1.000 & Complete \\
L4-RDT-009 & R & 1.000 & Unscored & Timeout \\
L4-RTO-001 & P & [0.490, 0.790] & [0.700, 1.000] & Complete \\
L4-RTO-001 & R & [0.520, 0.820] & [0.700, 1.000] & Complete \\
L4-RTO-001 & Z & [0.520, 0.820] & [0.700, 1.000] & Complete \\
L4-RTO-003 & R & [0.700, 1.000] & [0.700, 1.000] & Complete \\
}

\newcommand{\RFQUpdatedRows}{%
Original (amended) & 20 & 17/20 & 18/20\\
Additional (paired) & 19 & 19/19 & 16/19\\
Combined (paired) & 39 & 36/39 & 34/39\\
Combined (strict) & 36 & 33/36 & 32/36\\
}
\newcommand{\RFQUpdatedCoverageRows}{%
Original / Astra & 20 & 75.5 & 97.9 & 17/20\\
Original / Claude & 20 & 81.2 & 98.6 & 18/20\\
Additional / Astra & 19 & 85.0 & 100.0 & 19/19\\
Additional / Claude & 19 & 78.9 & 97.7 & 16/19\\
Combined / Astra & 39 & 80.1 & 98.9 & 36/39\\
Combined / Claude & 39 & 80.1 & 98.2 & 34/39\\
Strict / Astra & 36 & 80.8 & 98.8 & 33/36\\
Strict / Claude & 36 & 80.0 & 98.4 & 32/36\\
}
\newcommand{\RoneUpdatedRows}{%
AR-005 & 16.0 & 68.0 & 100.0 & 100.0\\
L4-CAM-001 (R) & 52.0--82.0 & 70.0--100.0 & 68.5--98.5 & 55.0--100.0\\
L4-BID-001 (Z) & 64.0--94.0 & 40.0--70.0 & 60.0--100.0 & 60.0--100.0\\
L4-ECO-001 (R) & 52.0--82.0 & 70.0--100.0 & 70.0--100.0 & 70.0--100.0\\
TIQ-029 & 52.6 & 99.3 & 98.7 & 99.3\\
TIQ-100 & 71.9 & 58.4 & 74.1 & 98.9\\
TIQ-085 & 23.1 & 84.6 & 80.0 & 97.7\\
}

\hypersetup{pdftitle={DRAFT-Bench: Evaluating Frontier Agents Across the Professional Design Lifecycle},pdfauthor={Rong Pan; Jianqiang Wang}}
\title{DRAFT-Bench: Evaluating Frontier Agents\\Across the Professional Design Lifecycle}
\author{%
  Rong Pan\thanks{Work done during an internship at Curify.} \\
  University of Maryland, College Park
  \And
  Jianqiang Wang \\
  curify-AI
}
\date{}

\begin{document}
\maketitle
\suppressfloats[t]
\begin{abstract}
Can frontier general-purpose agents understand client intent, explore creative directions, execute revisions, and prepare production deliverables? We introduce DRAFT-Bench (Design Reliability Across Feedback and Tasks), a lifecycle-oriented evaluation of requirement elicitation, creative exploration, constrained editing, iterative revision, and production handoff. Its Requirement-to-Artifact framework links evolving requirements, protected properties, artifact versions, and delivery obligations. We compare native Codex/Astra and Claude Code across 35 design briefs under 43 conditions, alongside a separate 40-scenario RFQ track. Astra completes 43/43 design conditions and Claude 35/43. All scored core runs receive full structural credit, yet on the 28 core conditions scored for both systems, semantic checks flag violations in 11 for Astra and 5 for Claude; these are model judgments, not expert-confirmed failure rates. RFQ coverage is high, but success ordering changes across cohorts. SKU expansion, revision dependencies, and White ink files expose ambiguity among unmet obligations, missing inputs, and evaluator interpretations. These findings separate execution, requirement compliance, and professional acceptance. Unpaired tracks and different tools preclude claims of RFQ-to-delivery success, factory acceptance, or a pure model ranking.
\end{abstract}

\begin{figure}[t]
\centering
\includegraphics[width=\textwidth]{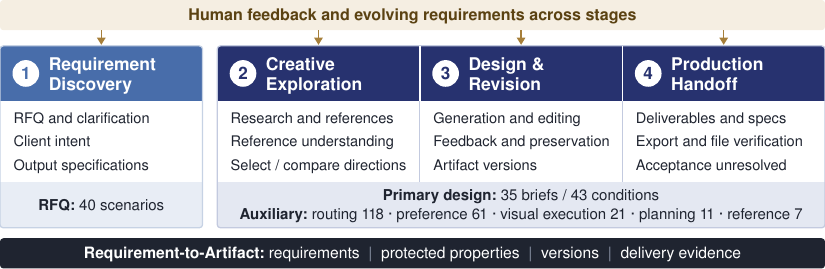}
\caption{Professional design lifecycle and DRAFT-Bench coverage. Human feedback can change requirements at every stage. Stages are evaluated through separate task protocols; RFQ and design execution are unpaired, and full RFQ-to-delivery execution is not yet tested.}
\label{fig:lifecycle}
\end{figure}

\section{Introduction}
A professional design assignment spans ambiguous client intent, creative exploration, constrained editing, iterative feedback, and preparation of deliverables for production handoff. General-purpose agents can participate in each stage, but a polished preview alone does not show whether they understood the request, preserved approved work, or delivered the required files. We ask: \emph{can frontier general-purpose agents translate evolving requirements into verifiable design artifacts across this lifecycle?}

We introduce \textbf{DRAFT-Bench}, \emph{Design Reliability Across Feedback and Tasks}, with 35 design briefs under 43 execution conditions and a separate 40-scenario requirement-elicitation track. Figure~\ref{fig:lifecycle} organizes the coverage from requirement discovery through production handoff. Creative exploration includes reference understanding, direction selection, and proposal comparison, alongside visual generation. The assignments are synthetic or curated and motivated by professional workflows; they are not a customer deployment study. \emph{Lifecycle-wide} describes coverage through separate protocols: RFQ and design execution are unpaired, so the study does not measure an integrated RFQ-to-delivery success rate. Broader archive accounting is reported separately in Section~\ref{sec:data}.

Our central methodological contribution is \textbf{Requirement-to-Artifact evaluation}: linking an active requirement to the relevant artifact version and the evidence needed to judge it. Requirements specify changes, protected properties, or deliverables. Feedback can activate an earlier obligation: choosing a packaging direction can trigger expansion into five SKUs. The framework makes this scope decision explicit before interpreting runtime status, candidate declarations, file facts, semantic judgments, or professional acceptance.

The configured systems show high intake coverage and substantial design execution capability, but differ in completion, delivered sets, and revision behavior. Structural credit coexists with semantic flags; client dependencies and inconsistent applicability decisions complicate interpretation. These observations can implicate the agent, the evaluator, or the specification. Our analysis preserves those distinctions instead of treating every discrepancy as a confirmed agent failure.

\paragraph{Research questions.}\mbox{}
\begin{introlist}
\item \textbf{RQ1: Lifecycle capabilities.} What do the separate protocols reveal about agents' capabilities and execution reliability across design stages?
\item \textbf{RQ2: Requirement-compliant delivery.} Do completed runs and structurally valid outputs satisfy active requirements?
\item \textbf{RQ3: Decisions and dependencies.} How do human choices, revision prerequisites, and production constraints affect delivery and its evaluation?
\end{introlist}

\paragraph{Contributions.}\mbox{}
\begin{introlist}
\item \textbf{Lifecycle coverage.} A task collection spans major stages of professional design, with briefs, assets, context conditions, and feedback motivated by production workflows. Tasks and toolkit: \url{https://github.com/super-inbox/DRAFT-Bench}.
\item \textbf{Requirement-to-Artifact evaluation.} An evidence framework links evolving requirements, protected properties, artifact versions, and delivery obligations. Its implementation combines an exhaustive audit of scored core checks with eight selected paired inspections; it is not a fully automated or expert-validated acceptance metric.
\item \textbf{Frontier agent capability analysis.} A comparison of Codex and Claude Code reveals differences between execution completion, semantic compliance signals, and evidence of usable delivery. Original scores remain frozen. Separate protocols and selected cases support stage-specific findings, not a universal system ranking or estimates of expert-accepted delivery.
\end{introlist}

\section{Related work and research gap}
\label{sec:related}
\paragraph{Design planning and task quality.}
GraphicBench couples planning to execution and already shows that plausible plans can fail in execution~\citep{ki2025graphicbench}. Its execution metrics include source-image fidelity and query-aligned visual questions. Graphic-Design-Bench spans layout, typography, infographics, template and design semantics, and animation, including structured output and editing~\citep{deganutti2026gdb}. These are not merely aesthetic leaderboards. Our emphasis is the evidence linking a changing delivery obligation to the correct version and the interpretation of disagreement between checks.

\paragraph{Interactive professional work.}
DeskCraft is the closest workflow comparison: it includes 538 tasks, mid-turn clarification and post-turn feedback, and domain-aware verifiers that extract state from project files or application runtimes~\citep{wang2026deskcraft}. It already evaluates professional delivery and human--agent collaboration. Proactive T2I agents connect clarification to image alignment and include human studies~\citep{hahn2025proactive}. Our RFQ experiment instead isolates attribution under fixed replies; its downstream connection is still untested. Drift-Bench++ studies evolving intent and imperfect communication through executable tasks and an interaction protocol~\citep{zhang2026drift}. Neither interactivity, intent tracking, nor artifact-based verification alone is a novelty claim.

\begin{table}[t]
\centering\small\setlength{\tabcolsep}{5pt}
\caption{Positioning by evaluation object. The final column states this paper's additional question, not an assertion that prior work lacks all related capabilities.}
\label{tab:related}
\sbox{\relatedworktable}{%
\begin{tabularx}{\textwidth}{@{}P{0.18\textwidth}YY@{}}
\toprule Work & Established evaluation object & \textbf{Additional question pursued here}\\\midrule
GraphicBench~\citep{ki2025graphicbench} & Plans, tool execution, source fidelity, and query alignment & Which current deliverable remains unresolved after a completed run?\\
Graphic-Design-Bench~\citep{deganutti2026gdb} & Design understanding, generation, structure, and semantics & Do valid outputs satisfy the latest workflow-specific obligations?\\
DeskCraft~\citep{wang2026deskcraft} & Professional desktop workflows, interaction, and state verification & How do design-specific constraints, declared blockers, and archived judgments disagree?\\
Drift-Bench++~\citep{zhang2026drift} & Interactive intent alignment under miscommunication and evolving goals & Which approved properties and deliverable versions substantiate the current design obligation?\\
TASTE; Agentic-DRS~\citep{zhu2026taste,nag2025agenticdrs} & Designer preferences; design diagnosis and actionable feedback & Was a requested change actually applied to the delivered version?\\
GDPval; industry agent-eval guidance~\citep{patwardhan2026gdpval,anthropic2026evals} & Expert evaluation of professional outputs; outcome and grader validity & Which design artifact and version substantiate each acceptance claim?\\
This study & Lifecycle-wide coverage through separate RFQ, exploration, revision, and handoff protocols & \textbf{Link evolving requirements, protected properties, and artifact versions to evidence of delivery.}\\\bottomrule
\end{tabularx}%
}%
\leavevmode\rlap{%
  \hspace{\dimexpr.59\textwidth+\tabcolsep\relax}%
  \color{questioncolumn}%
  \rule[-\dp\relatedworktable]%
    {\dimexpr.41\textwidth-\tabcolsep\relax}%
    {\dimexpr\ht\relatedworktable+\dp\relatedworktable\relax}%
}%
\usebox{\relatedworktable}
\end{table}

\paragraph{Judgment and professional acceptance.}
TASTE measures multidimensional designer preferences and evaluates model judges against professional panels~\citep{zhu2026taste}. Agentic-DRS studies design diagnosis and feedback quality with human comparisons~\citep{nag2025agenticdrs}. GDPval likewise uses occupational experts to assess professional deliverables~\citep{patwardhan2026gdpval}; Anthropic's evaluation guidance distinguishes traces from outcomes and stresses grader calibration~\citep{anthropic2026evals}. We build on these principles, rather than claiming that requirements or expert acceptance have been ignored.

\paragraph{Specific research gap.}
Prior work already covers planning, graphic quality, preferences, and professional workflows. DRAFT-Bench connects these concerns through lifecycle-wide design tasks and \emph{Requirement-to-Artifact evaluation}: which evolving obligation applies, which property must remain protected, and which artifact version substantiates delivery? This linkage supports a capability analysis that distinguishes completed execution, requirement compliance, and usable handoff, while exposing disagreement between evidence layers. We implement it through condition-level audits and selected paired inspections. Human acceptance calibration, an integrated RFQ-to-delivery study, and factory acceptance remain future work; neither professional workflow evaluation nor artifact verification alone is a novelty claim.

\section{DRAFT-Bench}
\label{sec:data}
\subsection{Archive scope and counting units}
The evaluation archive reconciles 232 design-related case lineages with 40 RFQ scenarios, yielding 272 lineages and 301 protocol units. Its 333 observation records retain 32 historical associations between the Brief Bank v0.2 core and v0.3; those associations add no new executions (Table~\ref{tab:scope}). Counts are not multiplied by the number of candidate systems. Routing, preference prediction, planning, and artifact execution measure different endpoints and cannot be combined into a common quality score. The 43-condition Brief Bank remains the primary design execution analysis.

\begin{table}[t]
\centering\small\setlength{\tabcolsep}{3pt}
\caption{Archive accounting. Lineages deduplicate IDs and versions; neither lineages nor protocol units establish independent statistical samples.}
\label{tab:scope}
\begin{tabularx}{\linewidth}{@{}Yrrr@{}}
\toprule Scope & Lineages & Units & Records\\\midrule
Design-related archive & 232 & 261 & 293\\
RFQ, two cohorts & 40 & 40 & 40\\\midrule
Combined & 272 & 301 & 333\\\bottomrule
\end{tabularx}
\end{table}

\subsection{Task structure and capability coverage}
A design condition specifies a brief, available source/reference assets, context, ordered feedback, constraints, and expected deliverables. The agent produces versioned artifacts and a response after each turn. The brief may contain obligations that become active only after a client choice; approved properties can remain fixed through later edits. This structure motivates checking the assignment across versions rather than judging only its last preview.

Figure~\ref{fig:lifecycle} groups these tasks by lifecycle stage. Requirement discovery uses RFQ fields, targeted questions, and fixed customer replies. Creative exploration includes alternative directions, reference understanding, direction selection, and comparison. Constrained edits and iterative revisions combine requested changes with protected content and version dependencies. Production handoff concerns complete sets, required formats, file structure, and declared handoff state. Table~\ref{tab:capability-map} maps this coverage to the conclusions supported by the current results.

\subsection{Primary design track}
Brief Bank data v0.3 contains 35 briefs under 43 conditions: 24 core briefs/32 conditions and eleven external briefs/conditions. Eight families cover identity, campaigns, packaging, e-commerce, format adaptation, feedback revision, reference-based work, and production handoff (Appendix~\ref{app:auxiliary}). Four core briefs each have zero-shot, reference-grounded, and personalized contexts. These conditions are not independent client projects. External tasks have newly authored assets and contracts inspired by public task patterns; they form a curated sample rather than an untouched population test.

Scripted feedback is delivered in order. It can refer to earlier choices or versions whether or not the candidate has established them. The version-dependency case in Section~\ref{sec:revision} shows why the feasibility of a feedback instruction must itself be inspected. We do not reinterpret a missing prerequisite as an agent failure without adjudication.

\subsection{Requirement-discovery track}
The original twenty synthetic English RFQ scenarios derive from communication patterns in five supplied examples and cover bulk images, tourist videos, cultural merchandise, content workflows, and apparel retouching. An additional twenty scenarios extend four supplied real-world scenario families: hotel-listing retouching, scene-enhancement pairs, educational animation, and a ski-opening poster. They retain authored variants and synthetic assumptions; the executed dialogues are simulations rather than real customer conversations. Each scenario has seven fields, four critical, and at most eight customer replies. Fixed answers expose information only when a corresponding question is asked; attachment mentions are placeholders.

The RFQ track measures requirement elicitation, not downstream design success. It is unpaired with the design track and cannot establish that additional coverage improves delivery. The two RFQ cohorts differ in auxiliary models, execution dates, and attempt histories; their combined view is a descriptive reconciliation, not a uniform new forty-case experiment. Auxiliary routing, preference, planning, and 21q results remain in Appendix~\ref{app:auxiliary}; overlapping observations are not added to the primary sample.

\section{Requirement-to-Artifact evaluation}
\label{sec:evaluation}
\subsection{Linking requirements, versions, and delivery evidence}
\begin{figure}[t]
\centering
\includegraphics[width=\textwidth]{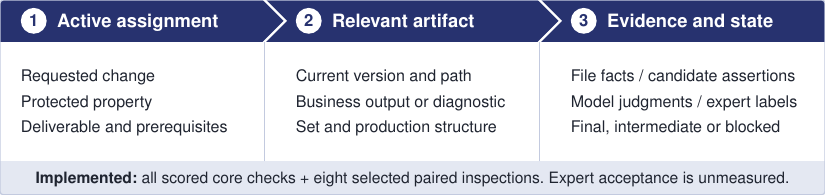}
\caption{Requirement-to-Artifact evidence framework. Applicability is checked before the requested version and its evidence. This is the audit schema; the archive supplies aggregate checks and selected direct inspections, not a fully automated acceptance system or an evaluated RFQ-to-delivery pipeline.}
\label{fig:protocol}
\end{figure}

The core evaluation unit is an active requirement paired with a relevant artifact version. Each record links the obligation to its originating turn, scope (requested change, protected property, or deliverable), artifact version/path, check result, and evidence provenance. Figure~\ref{fig:protocol} operationalizes the lifecycle view in Figure~\ref{fig:lifecycle} for the design-artifact audit. Evaluation proceeds through four questions:
\begin{enumerate}
\item \textbf{Is the requirement active?} Resolve selections, approvals, superseded requests, and unmet prerequisites. Exclude a requirement only with an explicit applicability reason.
\item \textbf{Which artifact is relevant?} Identify the requested version and business deliverable. A candidate-authored diagnostic or an older draft is not automatically its substitute.
\item \textbf{What changed and what was preserved?} Check the requested delta, protected content, complete set, and required file structure together.
\item \textbf{What supports the decision?} Distinguish observed file facts, candidate assertions, model judgments, and expert acceptance. Record unresolved dependencies and the candidate's declared status separately.
\end{enumerate}

\subsection{Three evidence layers}
\paragraph{Structural evidence.}
Check presence, readability, count, dimensions, format, and version identity. The archived \texttt{artifact\_contract} score is one such signal. Runtime completion is an execution property reported alongside it.

\paragraph{Semantic evidence.}
Check the task-specific change, source/reference roles, preservation constraints, and deliverable obligations. Unchanged pixels may indicate successful preservation or a missing edit; file existence alone cannot establish correctness. Archived rubric weights and scores are retained.

\paragraph{Workflow and handoff evidence.}
Distinguish final output, intermediate progress, and a declared client dependency. A blocker disclosure establishes that the candidate is waiting; it does not establish whether waiting was justified. Likewise, an unresolved delivery does not establish a false completion claim. Independent professional acceptance remains unmeasured.

\subsection{Implemented audit and its coverage}
For every scored core condition, we collect the archived semantic check list and flag the condition if any check has the exact verdict \texttt{VIOLATED}. Each condition counts once, independent of duplicated or differently worded checks. Coverage is complete over 32 Astra and 28 Claude scored core conditions; we also report the 28-condition intersection. No flag is not an acceptance certificate.

Direct inspection covers eight paired conditions: MFA-003 and CFR-002 from an initial audit round, plus PSF-001, RTO-001, ECO-001, BID-001, CFRY-003, and ECO-002. We inspect prompts, final responses, inventories, selected images and file internals. The six additional cases retain 23 original images and 76 source hashes. They are purposively selected to illustrate delivery transitions, evaluator ambiguity, and a positive retouching contrast; they do not estimate prevalence.

Historical altered-artifact controls address a separate evaluator question: whether stale candidate-authored evidence can override missing or unchanged outputs. Their retained summaries are not newly reproduced trials. The ECO-001 applicability comparison is likewise a diagnostic requiring expert review. We do not change the frozen judgments to favor an interpretation.

\paragraph{What the archived judge can observe.}
The main core scorer receives the condition-specific request, scripted feedback, constraints, deliverable requirements, input images, selected output images across version folders, business-deliverable text, and evaluator-computed file facts. It does not ingest the complete tool trace or the turn-final responses; candidate verification reports are excluded as proof. Output image selection defaults to at most ten images for core and twelve for external conditions, with further size limits and text truncation. The external scorer also receives aggregate trajectory counts. Expected workflow instructions are not a record of every action actually taken. Our direct case audit additionally reads final responses to distinguish disclosed dependencies from completion claims. This richer audit view does not retroactively change the judge's input or scores.

\paragraph{Separate rejudging pilot.}
The archive also retains R1, a seven-pair historical Sol/Astra pilot using revised task-specific acceptance rules and two anonymized, reversed-order presentations per pair. Its fourteen Gemini calls are distinct from our eight Astra/Claude artifact inspections. It changes neither candidate artifacts nor historical aggregate scores, excludes Claude, and has no final expert adjudication. Section~\ref{sec:grader} reports its diagnostic implications; Appendix~\ref{app:r1} preserves its scope.

\subsection{Endpoints and interpretation}
Reported endpoints are runtime completion, latency, archived quality bounds, and semantic-check flags; RFQ coverage and success are auxiliary. Timeouts leave quality unscored. Unmeasured rubric dimensions induce bounds, not confidence intervals. Quality combines task-specific model judgments and is not a human aesthetic rating.

An accepted-delivery rate requires independent acceptance labels. A false-completion-claim rate additionally requires an actual completion claim. Neither rate is available. Clarification usefulness should be evaluated separately while the corresponding deliverable remains unresolved. Appendix~\ref{app:protocol} supplies the metric formulas and inclusion rules.

\section{Experimental setup}
\label{sec:setup}
\paragraph{Design systems and budget.}
Evaluation v0.4 runs Codex CLI 0.154.0 with \model{gpt-6-astra} (three workers) and Claude Code 2.1.286 with \model{claude-opus-5-5} (two Brief Bank workers), both at max effort. Each attempts the same 43 conditions, with a 60-minute active-call limit \emph{per turn}, and receives scripted feedback in order. This is not a 60-minute total-project budget. Auxiliary 21q visual-execution cases have a 15-minute limit. Data and asset-manifest hashes match. Prompts, feedback ordering, weights, and judges are fixed. Subscription quota pauses precede 22 Claude turns and total about 71 hours across workers; this pacing can also change task order. Claude has no infrastructure retries in this design track; its eight failures are candidate timeouts. Historical Sol executions remain separate.

\paragraph{Tools and judges.}
Astra has Codex built-in image generation, bundled runtime packages, Codex skills, and MCP REPL tools. Claude uses native tools and code-based visual production without an image-generation model, MCP, skills, or subagents. Shell network access differs: Astra's three package installs fail on DNS inside its sandbox, while Claude installs packages in the one condition that tries. Neither system runs a web search. Claude Code sends images above 512,000 bytes as JPEG (quality 90 or lower) at original resolution without transparency (63 of 71 Brief Bank attachments, in 36 of 43 conditions); originals stay in each task's input folder, and no equivalent capture exists for Astra. Claude's repository isolation is enforced by the operating system, while Astra uses instruction-based read isolation. These are configured-system comparisons: model, runtime, platform instructions, tools, isolation, and date are confounded.

Gemini 2.5 Pro at temperature zero supplies archived artifact-grounded judgments (condition-aware v3 for core; v0.3 for external tasks). Under \texttt{v04-weighted-only-v1}, hard-gate vetoes are disabled and external score pass means weighted score $\geq0.70$. This archive-wide policy was adopted after Astra's scoring and before Claude's run, with no judge rerun. In the Brief Bank it changes one outcome: Astra's L4-RDT-007, at exactly 0.700, is no longer vetoed for a missing vector master, raising Astra's external passes from 10/11 to 11/11. Core full-score bounds preserve unmeasured dimensions. Neither a weighted pass nor a file-format check implies expert acceptance.

\paragraph{Analysis scope.}
We retain the design snapshot, its retrospective core-check audit, and eight paired direct case inspections. These are reconciled with the evaluation archive, whose source commit and released input hashes are in the public provenance file. The analysis adds no candidate or judge calls to these archived records. The evidence includes archived RFQ extension runs and R1 judge calls, but no completed expert study. There is one included execution per design condition; uncertainty across repeated candidate trials is unavailable. Per-turn durations exclude subscription waits before the turn. All completed-subset comparisons report their denominator and the omitted execution outcomes.

\paragraph{Auxiliary RFQ setup.}
The original RFQ cohort compares Codex/Astra, Codex/Sol and Claude Code/Opus under matched fixed-reply data and controller settings, with candidate business tools disabled. The additional cohort has scored Astra and Claude trajectories; Sol's subsequent attempt has no scoreable cases because of service rate limits. Router/auditor models and continuation histories differ across cohorts. Appendix~\ref{app:rfq} reports these settings, exclusions, paired denominators, and the original post-hoc amendment.

\section{Results: capability map and three findings}
\label{sec:results}
Table~\ref{tab:capability-map} maps the available evidence to the lifecycle in Figure~\ref{fig:lifecycle}. Coverage is assessed through separate protocols, not a single execution chain. We organize the results around stage-specific capabilities, requirement-compliant delivery, and the decisions that complicate evaluation.

\begin{table}[t]
\centering\small\setlength{\tabcolsep}{5pt}
\caption{Frontier agent capability map. Each row states the evidence available at one stage and its inference boundary; the rows are not a common quality scale or a measured RFQ-to-delivery trajectory.}
\label{tab:capability-map}
\begin{tabularx}{\textwidth}{@{}P{0.18\textwidth}P{0.27\textwidth}Y@{}}
\toprule Lifecycle stage & Existing evaluation & Supported conclusion and boundary\\\midrule
Requirement discovery & 40 RFQ scenarios in two cohorts & Both systems achieve high field coverage; success ordering changes across cohorts. Downstream benefit is untested.\\
Creative exploration & Direction, reference-understanding, planning, and preference tasks & Relevant capabilities are tested separately; their scores do not establish creative quality or real-client acceptance.\\
Design and revision & 35-brief / 43-condition Brief Bank; constrained edits and ordered feedback & Completion coexists with incomplete modifications and unresolved constraint or version dependencies.\\
Production handoff & Packaging sets, production PDFs, and file-structure checks & Output-set and file differences are observable; factory acceptance and manufacturability are untested.\\\bottomrule
\end{tabularx}
\end{table}

\subsection{Finding 1: broad capabilities, uneven execution reliability}
\label{sec:profiles}
The separate protocols demonstrate capabilities across design stages, while execution reliability differs between the configured systems. These are system-level observations under different tools and runtimes, not a pure model ranking.

\paragraph{Design execution, quality, and time.}
Astra completes all 43 design conditions; Claude completes 35/43 (81.4\%), with eight first-turn timeouts at the 60-minute limit. Four timeouts are core and four external. On the common 28 scored core conditions, Claude's observed weighted contribution is higher by \PairedCoreDelta{}, but full-score bounds are \PairedClaudeCore{} for Claude and \PairedAstraCore{} for Astra. The unconstrained difference range, \PairedCoreDifference{}, includes either ordering. These scores incorporate task performance; they do not isolate visual beauty.

For the seven shared external conditions, mean scores are 1.000 for Claude and \PairedAstraExternal{} for Astra, and both pass all seven score thresholds under the weighted-only rule; Astra's L4-RDT-007 sits exactly at 0.70 (Section~\ref{sec:setup}). On the planned external set, completion is 7/11 for Claude and 11/11 for Astra. Comparing only scored outputs selects on system behavior and can favor the system with more missing outcomes.

Across all 43 first turns, median recorded duration is \ClaudeFirstTurnMinutes{} minutes for Claude versus \AstraFirstTurnMinutes{} for Astra; the Claude median includes observations ending at the timeout deadline. Among completed turns, medians are \ClaudeCompletedTurnMinutes{} (77 turns) and \AstraCompletedTurnMinutes{} (90 turns). The eight timeouts are spread across six of eight task families (Appendix~\ref{app:category}); the small counts do not identify category-specific causal deficits.

\paragraph{Inference and open hypothesis.}
The systems occupy different observed completion--quality--latency profiles. A causal quality--reliability trade-off remains a hypothesis: image tools and other environment factors differ, both systems use max effort, and neither time budgets nor effort levels were experimentally varied. Additional time may or may not resolve the failures.

The broader archive reinforces the need to separate endpoints: Astra completes 259/261 protocol units and Claude 233/261, but these totals mix routing, planning, preference prediction, and artifact tasks. On 21q visual execution, completion under the 15-minute limit is 20/21 versus 2/21, with median durations of 2.07 and 15.01 minutes, while four Claude outputs are scoreable because two timed-out runs leave artifacts. On shared routing and preference tasks, differences are smaller or tied (Appendix~\ref{app:auxiliary}). These task profiles do not establish that planning quality transfers to execution, and their heterogeneous scores are not averaged.

\begin{figure}[t]
\centering
\includegraphics[width=\textwidth]{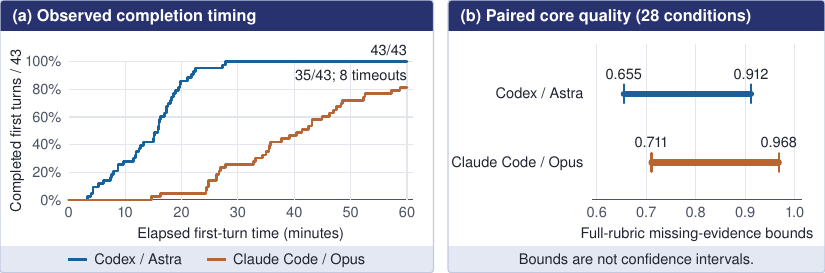}
\caption{Joint completion, timing, and quality profile. Left: cumulative first-turn completions in the recorded runs, among 43 planned conditions. Right: missing-evidence score bounds on the common 28 core conditions, excluding four Claude core timeouts. These are neither confidence intervals nor a controlled budget sweep. The panels avoid collapsing censored execution and incomplete quality evidence into a single rank.}
\label{fig:completion-quality}
\end{figure}

\paragraph{Creative exploration has distinct endpoints.}
Direction generation and reference use appear in the Brief Bank; auxiliary reference-planning tasks (scores 87.92 for Astra and 95.83 for Claude over six cases) and preference tasks (20/60 hits for each) assess interpretation and choice. These endpoints extend exploration beyond producing multiple images. They do not establish that a proposal is original, professionally strong, or accepted by a real client (Appendix~\ref{app:auxiliary}).

\paragraph{Requirement discovery: high coverage, cohort-sensitive ordering.}
\label{sec:rfq-results}
The expanded RFQ archive contains forty scenarios. On the original amended twenty, Claude has 18/20 successes, Astra 17/20, and Sol 11/20. On the additional nineteen jointly scored cases, Astra has 19/19 and Claude 16/19; Claude's infrastructure-excluded 028 is omitted from both sides. The small original ordering therefore reverses in the additional cohort (Table~\ref{tab:rfq-updated}). The combined paired view is 36/39 for Astra and 34/39 for Claude; removing original reattempt cases 004/009/019 gives 33/36 versus 32/36. The latter is an exclusion sensitivity check, not a correction for every cross-cohort difference.

\begin{table}[t]
\centering\small\setlength{\tabcolsep}{3pt}
\caption{RFQ success at eight replies on identical case IDs. Original means AMENDMENT-01. Strict excludes its three reattempt IDs as well as 028. Cohorts differ in auxiliary models and attempt histories.}
\label{tab:rfq-updated}
\begin{tabularx}{\linewidth}{@{}Yrrr@{}}
\toprule Cohort & $n$ & Astra & Claude\\\midrule
\RFQUpdatedRows
\bottomrule
\end{tabularx}
\end{table}

Combined elicited coverage at eight replies is 98.9\% for Astra and 98.2\% for Claude. High average coverage can coexist with missing a critical deliverable field: extension cases 022, 023, and 029 are archived Claude failures involving repeated or multi-field questions, routing ambiguity, or an early stop while output specifications remain unresolved. These are simulator/controller-dependent observations awaiting human review, not evidence of a universal questioning deficit. Among the 39 paired cases there are 32 joint successes, four Astra-only successes, two Claude-only successes, and one joint failure.

Appendix~\ref{app:rfq} retains cohort-specific settings, coverage and exclusions. Sol's additional twenty-case attempt has zero scoreable results because of HTTP 429 failures; it contributes no zero competence scores. Since RFQ and design tasks are unpaired, neither high coverage nor this success ordering establishes downstream design acceptance. That requires a separate execution experiment.

\subsection{Finding 2: completion does not guarantee compliant delivery}
\label{sec:delivery}
\paragraph{Observed cross-layer disagreement.}
All scored core runs receive a structural artifact-contract score of 1.0. Nevertheless, archived semantic checks flag at least one violation in \AstraCoreFlags{} Astra and \ClaudeCoreFlags{} Claude conditions (Table~\ref{tab:gap}). On the shared 28 core conditions, the counts are \AstraPairedCoreFlags{} and \ClaudePairedCoreFlags{}. Thus, structural credit and completed execution do not imply requirement-level compliance. These are retrospective model-judge flags, not independent expert failure labels; the smaller Claude population excludes four core timeouts.

\begin{table}[t]
\centering\small\setlength{\tabcolsep}{3pt}
\caption{Core evidence disagreement. Every scored core condition has a check list. A flag means any archived semantic check is \texttt{VIOLATED}; no flag is not an acceptance certificate.}
\label{tab:gap}
\begin{tabularx}{\linewidth}{@{}Yrr@{}}
\toprule Measure & Astra & Claude\\\midrule
Completed / attempted core & 32/32 & 28/32\\
Full structural credit / scored & 32/32 & 28/28\\
Any semantic flag / scored & 13/32 & 5/28\\
Any flag / shared core & 11/28 & 5/28\\\bottomrule
\end{tabularx}
\end{table}

\paragraph{A missing revision can be an explicit escalation.}
In MFA-003 both systems complete four runtime turns but export no requested business visual in v3. Both final responses explicitly state that the Story revision was not performed and request a client decision. In CFR-002, Astra's preview is byte-identical to the source; Claude changes pixels only within the title region while leaving the requested mark relocation unresolved. Both disclose the occupied target region as a blocker. Appendix~\ref{app:visual-cases} preserves the source/output comparison. These are unresolved deliveries; whether the pauses were professionally justified requires expert review of the request and assets.

\paragraph{Positive contrast: a bounded local edit.}
In ECO-002 both systems receive the same perfume-bottle image and export 1024$\times$1024 retouched PNGs. Both archived quality bounds are [0.900, 1.000], with no \texttt{VIOLATED} check. Appendix~\ref{app:visual-cases} shows the supplied image, outputs, and identical-coordinate crops. This supports a task-specific positive contrast to missing deliveries, not expert confirmation that every protected property was preserved. The relevant evaluation object combines the requested cleanup with constraints on the bottle, label, cap, color, and framing.

\subsection{Finding 3: decisions, revisions, and production constraints expose evaluation ambiguity}
\label{sec:revision}
Three cases illustrate why delivery must be evaluated against the active requirement and the relevant artifact version. A direction choice changes the requested set; scripted feedback can assume a missing prerequisite; and a file can contain a required resource without establishing production suitability. These are selected mechanisms, not estimates of failure prevalence.

\begin{figure}[t]
\centering
\includegraphics[width=\textwidth]{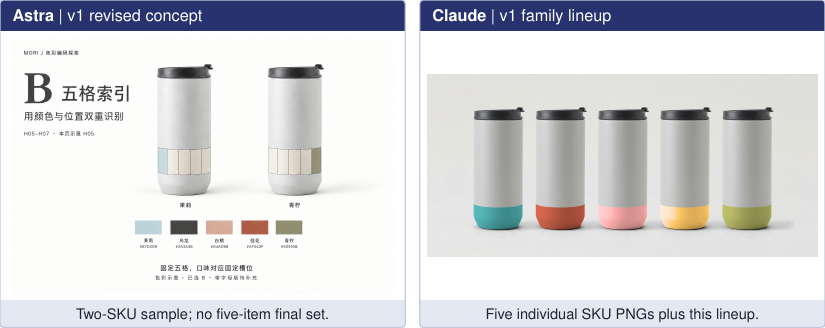}
\caption{PSF-001 after selection. Left: Astra's revised two-SKU sample. Right: Claude's five-SKU lineup, accompanied by five individual PNGs. Completing the local hue edit and expanding the selected strategy are distinct obligations. Original artifact pixels are preserved; professional acceptance remains unmeasured.}
\label{fig:sku}
\end{figure}

\paragraph{A selection changes the deliverable scope.}
In reference-grounded PSF-001, the brief requests five packaging SKUs after a strategy is chosen. Feedback selects strategy 2 and changes the lime hue. Astra updates a two-SKU concept sample; Claude exports five individual SKU PNGs and a family lineup, verified from the inventory (Figure~\ref{fig:sku}). A brief-level reading activates expansion after selection; the runner also instructs the agent to apply only the requested delta. Expert acceptance must reconcile these scopes before attributing an omission. Both systems still identify missing approved typography or production inputs. The comparison establishes different output sets, not factory acceptance or superior visual quality.

RTO-001 provides a related transition: after selection and a visual-language revision, Astra delivers one revised concept while Claude exports the requested hero and two modules. Claude also discloses ambiguity about the selected direction. Appendix~\ref{app:visual-cases} shows the outputs; their presence does not adjudicate originality or that ambiguity.

\begin{figure}[t]
\centering
\includegraphics[width=\textwidth]{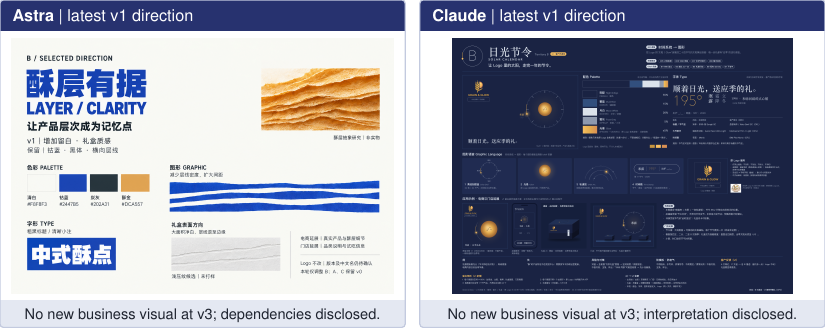}
\caption{BID-001: latest v1 direction visuals, shown because neither system has a new business visual at v3. Final responses explicitly disclose missing dependencies or an uncertain version interpretation. The later feedback's assumptions and the appropriateness of clarification need adjudication.}
\label{fig:version}
\end{figure}

\paragraph{Feedback can assume an unestablished version.}
BID-001 asks the agent to keep a v2 gift box and marketplace image while restoring a standee's whitespace to v1. Both complete four turns, but their last business visuals remain at v1; the v3 directories contain records or diagnostic evidence. Astra requests the missing standee files; Claude asks which store display the feedback refers to and explains its version interpretation. The script assumes prior artifacts or decisions that are not established in the outputs. Figure~\ref{fig:version} shows the latest business visuals. Evaluation must check this precondition before assigning blame to the candidate.

\begin{figure}[t]
\centering
\includegraphics[width=\textwidth]{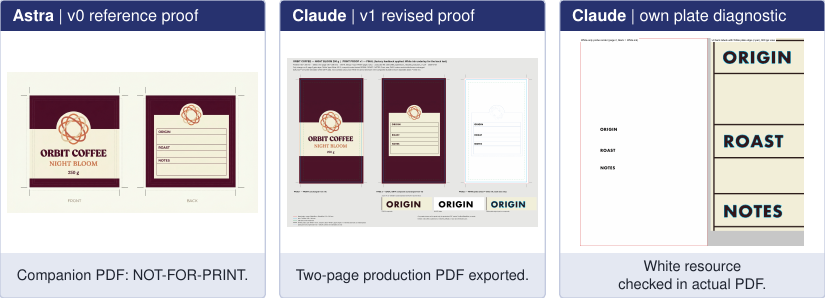}
\caption{CFRY-003. Astra's reference proof, Claude's revised proof, and Claude's own White-plate diagnostic. The diagram at right is candidate-authored evidence. We separately inspect the actual PDF for a White separation resource. Plate correctness and the text-color ambiguity remain unadjudicated; factory acceptance is not tested.}
\label{fig:production}
\end{figure}

\paragraph{Production handoff needs more than a proof.}
CFRY-003 requests a proof, a production PDF, and preflight, followed by a White ink layer under back text while other properties remain fixed. Astra provides a v0 reference proof and a PDF named \texttt{NOT-FOR-PRINT}, then requests an approved source rather than exporting a v1 production PDF. Claude exports a two-page v1 PDF. Direct inspection confirms a \texttt{/Separation} color-space resource named \texttt{/White}, beyond the candidate's diagnostic image (Figure~\ref{fig:production}). Both raise a mismatch between feedback about light text and dark text in the artwork. Resource presence does not establish correct plate placement, print suitability, or factory acceptance. The evaluated stage is production handoff; actual manufacturing acceptance remains future work.

\paragraph{Consequences for Requirement-to-Artifact evaluation.}
\label{sec:grader}
Across these cases, a missing output, a disclosed dependency, and insufficient acceptance evidence require different labels. The following evaluator diagnostics show why that distinction also depends on the assessment procedure.

\paragraph{Applicability can diverge after the same feedback.}
ECO-001 initially requests six launch assets after direction selection. Both systems receive identical feedback explicitly selecting the urban-commute direction and asking for quieter styling while preserving product color. Astra revises a direction board; its archived six-final-assets check is \texttt{NOT\_APPLICABLE}, with the reason that selection is still pending. Claude exports six images, but labels the hanging-loop panel \texttt{PROVISIONAL}: the supplied product view does not show that feature, and its final reply asks for another view. Claude's complete-set check is \texttt{VIOLATED} (Figure~\ref{fig:applicability}).

This case separates candidate behavior from evaluator behavior. Disclosure of a provisional asset is relevant evidence, but cannot by itself explain why the same selection feedback leads to different scope decisions. The archived R1 pilot below explicitly considers a narrower, current-delta interpretation for historical Sol/Astra outputs of this condition; it does not rejudge Claude or resolve which interpretation is professionally correct. The appropriate applicability and acceptance decisions remain unadjudicated. In particular, the absence of a semantic flag cannot be interpreted as accepted delivery.

\paragraph{Revised rules change judgments on unchanged artifacts.}
R1 rejudges seven selected historical Sol/Astra pairs, using task-specific rules and two anonymized presentations with reversed order per pair. AR-005's text-ranking reports change from 16/100 and 68/100 to 100/100 for both under a report-appropriate rubric. BID-001's original [64,94] versus [40,70] becomes [60,100] for both when unestablished feedback prerequisites are separated from observable revision quality. The upper bounds remain unknown contributions, not earned scores. Three pairs vary across the two presentations; Sol's TIQ-100 receives 59.1 and 89.1. Thus, rule/evidence changes can remove apparent differences without candidate improvement, but the revised judge is not validated as correct or stable. Order and judge variability are confounded. The pilot awaits expert adjudication and is kept outside all original aggregates (Appendix~\ref{app:r1}).

\begin{figure}[t]
\centering
\includegraphics[width=\linewidth]{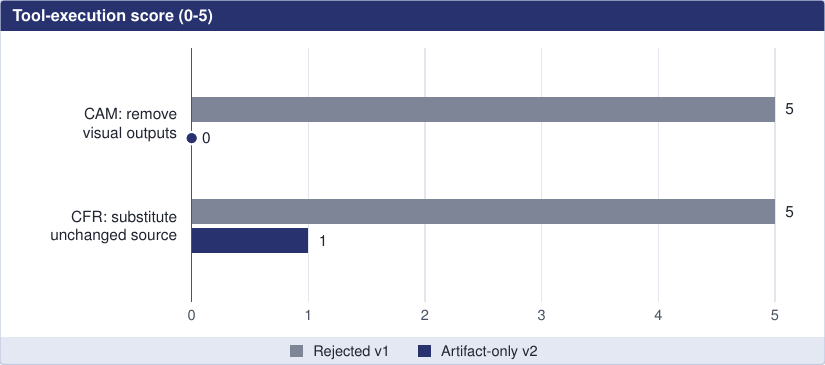}
\caption{Historical tool-judge controls, summarized from the retained validation document. The intervention changes artifacts while leaving stale self-evidence in place. Raw temporary v1 scores were discarded in the archive; the visual summarizes the recorded controls, not a newly reproduced experiment.}
\label{fig:controls}
\end{figure}

\paragraph{Historical stale-evidence controls.}
Historical tests kept candidate-authored success evidence while (i) deleting CAM business PNG/SVG outputs or (ii) replacing a CFR preview with the unchanged source. The first tool judge still assigned 5/5 to each. An artifact-only replacement, supplied with independently collected file facts and source/output images, assigned 0/5 and 1/5 (Figure~\ref{fig:controls}). A separate external no-op control reduced edit fidelity from 5/5 to 0/5 under its revised evidence view.

These interventions show sensitivity to a specific evidence-boundary defect. They do not isolate prose from every other candidate-authored metric, establish intentional reward hacking, or validate all judgments. The 21q archive contains the opposite risk: Claude's AR-001 output is called unchanged, while an archived same-size RGB audit finds 69,746 changed pixels out of 1,069,056 (6.52\%). This contradicts literal pixel identity, without certifying that the enlargement, preservation, or quality requirements were met. The frozen score remains unchanged. In AR-010, the archived judgment describes only raster delivery despite the inventory containing SVG, DXF, and PDF files; their existence warrants reinspection but does not establish production suitability.

\paragraph{Inference.}
A valid evaluation needs evidence about the evaluator as well as the agent. The historical altered-artifact controls and R1 rejudging pilot provide two distinct diagnostics: changing outputs while holding stale claims, and changing the assessment procedure while holding outputs. Neither estimates expert-calibrated sensitivity or false-positive rates. Wrong-version delivery and unintended-edit controls, followed by representative expert review, remain necessary.

\section{Discussion and limitations}
\label{sec:discussion}
\paragraph{Lifecycle coverage and delivery reliability.}
DRAFT-Bench provides evidence across requirement discovery, creative exploration, design and revision, and production handoff. Its Requirement-to-Artifact framework makes the linkage between evolving obligations and delivered versions explicit. The three findings separate stage-specific capability, requirement compliance, and the interpretation of missing or ambiguous evidence. Astra completes more of the observed execution workload, while Claude has a higher observed weighted contribution on the selected shared Brief Bank core. Missing outcomes, unknown rubric contributions, and different tools prevent a total quality ranking. Neither routing and planning scores nor completed execution establish independent professional acceptance.

\paragraph{Intake coverage is a separate capability.}
RFQ's near-complete average field coverage does not establish that every critical output requirement was elicited, that the final brief is usable, or that a downstream design will be accepted. The success ordering changes between the original and additional cohorts and narrows under the exclusion sensitivity check. Auxiliary models, continuation history, and scripted disclosure also change, so this is descriptive heterogeneity rather than an isolated domain effect. A claim that either system is generally better at elicitation exceeds these data.

\paragraph{The evaluator is part of the measurement.}
The unchanged-artifact R1 pilot demonstrates sensitivity to acceptance rules and evidence presentation: a text-ranking task need not require an image, and a missing edit may be an unfulfilled obligation, justified clarification, or an invalid feedback prerequisite. These states require different labels. R1's remaining variation and lack of expert adjudication prevent calling the new scores corrected ground truth. Its seven selected pairs cannot estimate a benchmark-wide grading error rate. The main scorer's image sampling and incomplete dialogue view add limits: evidence available in the archive is not necessarily evidence delivered to the judge.

\paragraph{Implications for agent and evaluator development.}
Track active obligations and approved properties across versions; make the current checkpoint and the eventual deliverable set explicit. Reconcile brief-level scope with turn-level delta instructions before judging an omission. Provide artifact facts independently of candidate claims, and retain the candidate's outward explanation as evidence of communication, not as proof that a visual edit occurred. Separately label delivery completeness, the reason for blocking, and whether escalation was justified. These are development targets motivated by the audit, not interventions validated by this study. They align with professional-workflow evaluation~\citep{openai2026ironclad} and guidance on outcomes and grader calibration~\citep{anthropic2026evals}.

\paragraph{The next discriminating experiments.}
First use disputed cases to specify acceptance rules, then freeze those rules and evaluate a separate, stratified sample with experts blinded to system identity, including flagged and unflagged outputs and blocker decisions. Second compare judges on the same frozen evidence and report agreement and errors against expert labels; keep the R1 development cases separate from that test. Third compare agents with matched tools and repeated runs, and connect vague, elicited, and complete briefs to one downstream executor. This would test RFQ's delivery value and enable validation of an integrated RFQ-to-handoff chain. Production experts should separately assess plate placement, preflight, and factory acceptance before making design-to-manufacturing claims. None of these experiments is completed here.

\paragraph{Scope and causal limits.}
The 272 lineages and 301 protocol units span curated, synthetic, and historical tasks; the 333 records include duplicate associations and are not independent observations. Design executions are single included trajectories. RFQ variants share source families, its original amendment is post-hoc, and the cohorts use different auxiliary models and attempt histories. Timeouts, infrastructure exclusions, and missing quality dimensions remain distinct. Tools, runtimes, platform instructions, isolation, and dates differ across systems, preventing model-weight attribution or a causal quality--reliability trade-off. Human acceptance calibration remains absent. Production handoff is evaluated through artifacts and file evidence; hashes, pixel differences, and a White separation resource cannot certify professional quality or factory acceptance. Some rejected historical judge outputs were discarded.

\section{Conclusion}
DRAFT-Bench evaluates frontier agents across the professional design lifecycle through separate task protocols. Its central contribution is Requirement-to-Artifact evaluation: linking evolving requirements, protected properties, artifact versions, and delivery obligations to inspectable evidence. Across 35 design briefs and 43 conditions, Astra completes 43/43 and Claude 35/43, while full structural credit coexists with semantic flags and unresolved deliveries. The separate 40-scenario RFQ track shows high coverage and cohort-sensitive success ordering; the shared-case totals are 36/39 for Astra and 34/39 for Claude. SKU expansion, version dependencies, and White ink files show why human decisions and production constraints must be part of evaluating delivery. These results support a stage-specific capability analysis, with distinct measures of execution, requirement compliance, and justified escalation. They do not establish a pure model ranking, expert-accepted delivery, or factory acceptance. A paired RFQ-to-handoff experiment and independent expert calibration are the next steps beyond the current lifecycle-wide benchmark.

\paragraph{Data and code availability.}
Public materials are at \url{https://github.com/super-inbox/DRAFT-Bench}. Data release \texttt{draftbench-v0.1.0} (CC BY 4.0) contains the briefs, 24 input fixtures, RFQ scenarios and 100 run records with offline score replay, numeric records for all 301 protocol units, and the R1 scores. Toolkit release \texttt{draftbench-v0.2.0} (new code under MIT) runs these tasks under a new protocol; it does not reproduce the historical environments. Task inputs for the auxiliary protocols, candidate outputs beyond our figures, complete traces, judge payloads, the semantic-check audit, case evidence, and this paper's analysis scripts are not public, so the design judgments cannot be re-created from these materials alone.

\bibliographystyle{plainnat}
\bibliography{references}

\clearpage
\appendix
\section{Auxiliary RFQ results and source selection}
\label{app:rfq}
\subsection{Original cohort: systems and execution policy}
We compare Codex/\model{gpt-6-astra}, Codex/\model{gpt-5.6-sol}, and Claude Code/\model{claude-opus-5-5}, all at max effort. Runtime versions are Codex 0.154.0 and Claude Code 2.1.286. Router and independent annotation roles use \model{gpt-5.6-sol} at medium and high effort, respectively. ``Independent'' denotes a fresh evaluation role, not a different model family or human adjudication. Customers use fixed replies, with no generation model. Frozen data, prompts, projections, controller, scorer, and router/auditor configuration match; candidate business tools are disabled.

Claude's original 17-case analysis excludes three infrastructure failures. A disclosed \emph{post-hoc} amendment includes the first reattempt of each, yielding twenty cases. Both results stay visible; the later successful diagnostic rerun of 019 is excluded. Appendix~\ref{app:protocol} gives the complete accounting.

\subsection{Original cohort results}
Amended Claude reaches \ClaudeRFQCoverageSix{} elicited coverage by six replies and \ClaudeRFQCoverageEight{} by eight, with 18/20 successes. Astra has 17/20 and Sol 11/20 (Table~\ref{tab:rfq-main}). Claude's original policy yields 16/17 with three infrastructure exclusions; the twenty-case result uses a post-hoc reattempt amendment. The additional success over Astra and the 0.7-point C@8 gap are small descriptive differences, not population-level superiority.

\begin{table}[t]
\centering\small\setlength{\tabcolsep}{3pt}
\caption{Original RFQ cohort: coverage (percent), success at eight replies, and leakage counts. Claude's amended set is post-hoc; the original row has a different denominator.}
\label{tab:rfq-main}
\begin{tabularx}{\linewidth}{@{}Yrrrrr@{}}
\toprule System & $n$ & C@6 & C@8 & S@8 & Leak\\\midrule
\RFQComparisonRows
\bottomrule
\end{tabularx}
\end{table}

\subsection{Additional cohort and combined accounting}
The additional RFQ-IC-021--040 cohort uses Claude Code 2.1.299 and Codex CLI 0.154.0 with the same candidate model identities and max effort. Its router and auditor use Astra at medium and high effort, whereas the original cohort uses Sol in those roles. Fixed customer replies are retained. The Astra reference continues an interrupted trajectory, preserving thirteen earlier candidate questions and one customer reply previously routed by Sol; Claude starts from the opening. These differences limit attribution to the candidate model.

Claude has 16 successes among nineteen scored cases, with case 028 excluded for an infrastructure failure; Astra succeeds on all twenty. The paired additional view excludes 028 for both systems and yields 19/19 versus 16/19. Combining with the original amended cohort gives 36/39 versus 34/39. The strict view removes 004, 009, 019, and 028 from both sides, yielding 33/36 versus 32/36. It removes the original Claude reattempts but does not erase the Astra continuation or auxiliary-model differences. Full-system views retain Astra's 37/40 and Claude's 34/39 scored out of forty planned. Only the equal-membership rows in Table~\ref{tab:rfq-coverage-updated} are paired comparisons.

\begin{table}[t]
\centering\small\setlength{\tabcolsep}{3pt}
\caption{RFQ cohort-specific coverage percentages and success. The original rows use AMENDMENT-01; combined and strict rows summarize heterogeneous archived runs.}
\label{tab:rfq-coverage-updated}
\begin{tabularx}{\linewidth}{@{}Yrrrr@{}}
\toprule Cohort / system & $n$ & C@6 & C@8 & S@8\\\midrule
\RFQUpdatedCoverageRows
\bottomrule
\end{tabularx}
\end{table}

Coverage is the mean of case metrics using each case's initially unknown field denominator. In the paired 39, unprompted simulator disclosures are 2/270 for Astra and 0/266 for Claude; these are disclosure-policy diagnostics, not separate candidate competence scores. Additional-cohort cases 022/023/029 illustrate unresolved output fields under the fixed-reply policy. Human review remains unperformed; automated final-brief diagnostics do not alter the success gate. The original 006/011 attribution disputes also remain unresolved.

A subsequent Sol attempt on the additional twenty cases records 64 native calls, including 60 HTTP 429 failures, only two completed customer rounds, and zero scoreable cases. All twenty are infrastructure exclusions. This attempt is reported separately from the combined RFQ view, where these Sol trajectories are unavailable; no zero scores or claims about current service availability are inferred.

\paragraph{Interpretation and missing experiment.}
Near-complete intake coverage does not measure the value of a usable brief. Our RFQ and execution tasks are unpaired, and known fields do not resolve all spatial or approval ambiguities during editing. A decisive next test supplies one executor with a vague brief, an agent-elicited brief, and a complete reference brief, holding assets and tools fixed. It should measure accepted delivery and justified clarification, not merely another coverage score. The current evidence supports separating these questions, not claiming a measured downstream benefit.

\subsection{Original-cohort paired cases}
Table~\ref{tab:rfq-paired} uses the common amended set. S/A/C denote Sol/Astra/Claude; Y/N denote success. Asterisks mark Claude's first infrastructure reattempts; other results use the formal first attempt. The later diagnostic 019 rerun is excluded.

\begin{table}[ht]
\centering\small\setlength{\tabcolsep}{4pt}
\caption{RFQ C@8 percentages and success on identical case IDs. Original Claude analysis excludes 004, 009, and 019.}
\label{tab:rfq-paired}
\begin{tabularx}{\linewidth}{@{}Yrrrr@{}}
\toprule Case & Sol & Astra & Claude & S/A/C\\\midrule
\RFQPairedCaseRows
\bottomrule
\end{tabularx}
\end{table}

\section{Historical design context}
\label{app:legacy}
The Codex/Sol archive contains 32 core conditions inherited from v0.2 and eleven new external conditions. Core bounds are [\CoreLower, \CoreUpper]; the external mean is \ExternalScore{}, with 11/11 passes under the \emph{historical hard-gate plus score} rule. Core execution was selected from 66 attempts (32 completed, 31 errors, three timeouts). These records cannot supply a first-attempt reliability baseline or be pooled with the v0.4 results. The historical ECO-002 retouching omission disappears in the judgments of both v0.4 systems, but protocol and system changes confound attribution of this improvement.

\section{Supporting archives}
\label{app:auxiliary}
The design-related archive contains 232 case lineages, 261 protocol execution units, and 293 records, including 32 duplicate core associations between Brief Bank v0.2 and v0.3. Counting Brief Bank by its 35 briefs, the six groups in Table~\ref{tab:archive-completion} hold 253 case IDs. All 21 IDs in 21q also appear in routing under a different protocol, so they count once toward the 232 lineages and twice toward the 261 units. Adding forty RFQ scenarios yields the 272/301/333 accounting in Table~\ref{tab:scope}. These counts are not numbers of independent client projects or successful deliveries.

Across the 261 design-related units, Astra has 259 completed and 260 scored runs; Claude has 233 completed and 235 scored runs. Completed status and scoreability differ because a failed or timed-out run can leave assessable artifacts. Table~\ref{tab:archive-completion} reports each protocol separately. Brief Bank v0.2 adds 32 historical associations, but no additional units.

\begin{table}[t]
\centering\small\setlength{\tabcolsep}{3pt}
\caption{Design-related archive completion. C/S means completed/scored; the counts are out of $n$ in each row. No common quality scale is implied.}
\label{tab:archive-completion}
\begin{tabularx}{\linewidth}{@{}Ylrrr@{}}
\toprule Protocol & Output & $n$ & Astra C/S & Claude C/S\\\midrule
Brief Bank v0.3 & Artifacts & 43 & 43/43 & 35/35\\
21q visual execution & Artifacts & 21 & 20/21 & 2/4\\
Routing & Route & 118 & 118/118 & 118/118\\
99designs preference & Ranking & 61 & 61/61 & 60/60\\
Workflow planning & Plan & 11 & 11/11 & 11/11\\
Zcool reference planning & Plan & 7 & 6/6 & 7/7\\\midrule
Total & & 261 & 259/260 & 233/235\\\bottomrule
\end{tabularx}
\end{table}

On shared cases, routing joint-pass counts are 73/100 for Astra and 75/100 for Claude, and agent-route selection is 16/18 for each. Customer-preference hits are 20/60 for each. Workflow planning scores are 67.01 and 68.66 over eleven cases; reference-planning scores are 87.92 and 95.83 over six common cases. Those are protocol-specific results rather than evidence of executed visual delivery. The 21q score means, 72.58 versus 61.17 on four shared cases, are strongly selected; two of those Claude runs time out and some judgments have evidence disputes. These slices are not pooled into the primary design quality result.

\paragraph{Task identifiers.}
Brief Bank IDs such as DAB-L3-CFR-002 give level, family, and number; shorter forms drop the prefix or level. L3 tasks are single-turn, with no scripted feedback; L4 tasks add one to three feedback turns, multiple deliverables, and a human checkpoint. Family codes are BID (brand identity), CAM (campaign), PSF (packaging), ECO (e-commerce), MFA (multi-format adaptation), CFR (client-feedback revision), RTO (reference-to-original), and CFRY (production handoff); RDT marks the eleven external tasks. Routing has 18 agent-route (AR-) and 100 tool-intent (TIQ-) queries; 21q executes 21 of these IDs (12 AR, 9 TIQ) as artifact tasks.

\section{Historical R1 rejudging pilot}
\label{app:r1}
R1 covers seven selected historical Sol/Astra pairs, not Claude, with fourteen Gemini 2.5 Pro calls at temperature zero: two anonymized, reversed-order presentations per pair. Revised rules evaluate only candidate-visible requirements, treat a text report as a valid output where appropriate, distinguish current-turn scope from future delivery, and leave revision quality unobserved when necessary clarification or an invalid prerequisite prevents evaluation. The detailed policy and every new score remain provisional, pending expert adjudication. Table~\ref{tab:r1} reports the pilot without replacing any historical scores or whole-benchmark means.

\begin{table}[t]
\centering\small\setlength{\tabcolsep}{7pt}
\caption{R1 scores on a 0--100 scale. R/Z denote reference-grounded/zero-shot. R1 averages the two presentations' bounds; intervals represent missing evidence, not confidence intervals or earned upper scores. TIQ-100's Sol mean masks a 30-point discrepancy.}
\label{tab:r1}
\begin{tabularx}{\textwidth}{@{}Yrrrr@{}}
\toprule Case & Sol original & Astra original & Sol R1 & Astra R1\\\midrule
\RoneUpdatedRows
\bottomrule
\end{tabularx}
\end{table}

In CAM-001, the pilot treats Astra's paused refinement as unobserved, widening its interval to [55,100] rather than awarding full refinement credit. BID-001 and ECO-001 become equal intervals across the two candidates under R1's stage interpretation, while remaining delivery gaps are retained. TIQ-100 remains unstable: Sol receives 59.1 and 89.1 despite unchanged artifacts, so 74.1 is only an arithmetic average. CAM-001 and TIQ-085 also vary. Two presentations cannot separate order effects from judge variability. Changes jointly involve rules, evidence presentation, and scoring; they do not isolate the causal effect of one prompt change. This selected pilot supplies no independent expert labels, general error rates, or revised Claude comparison.

\section{Validation and the next decisive experiments}
\label{app:todo}
The retained numerical analysis recomputes per-case RFQ metrics, verifies amendment source hashes, checks 86 design run records, and derives quality bounds and timing from condition and turn records. A final reconciliation additionally verifies ten archive source files against a pinned source commit, reproduces the 272/301/333 accounting, recomputes paired and strict RFQ summaries from case records, and checks the fourteen R1 repetitions. The public release supports offline recomputation of the archive accounting and RFQ summaries; the design snapshot and regeneration scripts remain in the project archive. Hashes identify inspected files, not hidden provider weights or semantic correctness.

The next experiments should target the remaining causal questions:
\begin{enumerate}
\item Match image tools and vary effort/time budgets within each system; measure independently accepted deliveries.
\item Blindly adjudicate delivery omissions, partial edits, ceiling scores, and the no-op dispute; test judges with altered artifacts.
\item Repeat held-out, brief-clustered evaluations with RFQ inclusion policies declared before execution.
\item Compare downstream execution from vague, elicited, and complete briefs with one executor and shared assets.
\item Compare adaptive RFQ intake with a fixed seven-field questionnaire under the same reply policy.
\end{enumerate}

\section{Metric definitions and protocol accounting}
\label{app:protocol}
\subsection{RFQ attribution}
An RFQ round is an agent message and customer reply. The opening is free; clarification, repetition, and confirmation consume replies. Let $F_i$ be the applicable fields unknown after the opening, and let $e_{if}(t)$ indicate the first complete answer to field $f$ elicited by an appropriate question by round $t$. Case and mean coverage are
\begin{equation}
 C_i(t)=\frac{\sum_{f\in F_i}e_{if}(t)}{|F_i|},\qquad
 \overline C(t)=\frac{1}{N}\sum_i C_i(t).
\end{equation}
For initially unknown critical fields $K_i$, success is
\begin{equation}
 S_i=\mathbf 1[C_i(8)\geq0.8]\prod_{f\in K_i}e_{if}(8).
\end{equation}
An early-ended episode carries its last state forward; a final summary cannot retroactively earn credit. A separate annotation pass checks attribution without rewriting the customer's actual reply. Unsolicited disclosure affects observed coverage, not elicited coverage. Leak rate is unsolicited new fields divided by all newly disclosed fields, deduplicated per case; it is not a separate candidate competence score.

In the original twenty-case cohort, eighteen openings leave seven fields unknown and two leave six (138 total). Because the policy releases at most one new field per reply, its ideal six-round macro ceiling is $(18\times6/7+2)/20=0.8714$. This is a protocol bound for that cohort, not an executed questionnaire baseline or a combined forty-case estimate.

\subsection{Post-hoc RFQ amendment}
The original Claude analysis excludes duplicate-field router failures on 004/009 and a subscription-limit failure on 019. AMENDMENT-01 was adopted after original and supplementary outcomes were known. It selects the first frozen-configuration reattempt for every excluded case, including the failed 019 reattempt. A later diagnostic success on 019 is never selected. On the common original 17 cases, success is 16/17 for Claude, 14/17 for Astra, and 9/17 for Sol.

Original verification retains 152 plugin-allowlist failures out of 3,653 checks; a separate addendum accepts those built-in plugins. The original record is not relabeled as wholly successful. Cases 006/011 favor amended Claude over Astra, but the earlier Astra failures involve attribution disputes. Case 008 fails for both; these details limit behavioral claims from the one-case total difference.

\subsection{Quality bounds and selection}
For applicable dimensions $A$, observed dimensions $O$, normalized scores $s_d$, and weights $w_d$, we retain
\begin{align}
 L &= \frac{\sum_{d\in O}w_ds_d}{\sum_{d\in A}w_d},\\
 U &= L+\frac{\sum_{d\in A\setminus O}w_d}{\sum_{d\in A}w_d}.
\end{align}
These are missing-evidence bounds. Workflow completion and recovery have instrumentation gaps; an observed omission is a different kind of evidence and may legitimately receive a low semantic score. The paired full-score difference is bounded by $[L_C-U_A,U_C-L_A]$. Unmeasured weight is equal within each paired core condition, but its unmeasured performance need not be equal.

The archive's summary comparison of 0.871 (Claude) versus 0.816 (Astra) mixes interval midpoints and external point scores over 35 selected pairs. Its exploratory sign test concerns that derived diagnostic. We do not use it as a measured total score or a population-level quality ranking. Several conditions share a brief, and multiple tests were uncorrected.

\subsection{Retrospective requirement-check audit}
For each scored core condition $i$, the descriptive flag is $V_i=1$ if at least one archived check is exactly \texttt{VIOLATED}. Empty check lists would be unresolved rather than negative flags; none occur in the 32/28 scored core records. We do not use the spelling of a deliverable label to identify violations, avoiding capitalization and wording differences. Counting conditions rather than checks prevents duplicated requirements from inflating the rate. The unflagged complement is not an acceptance rate.

The audit's inference is deliberately limited: a full structural score coexists with a model-judged semantic violation. Cases may contain judge mistakes, underspecified requests, or justified requests for clarification. Hash checks and pixel differences do not adjudicate those possibilities. All original scores remain unchanged.

\section{Execution failures by task family}
\label{app:category}
Table~\ref{tab:category} counts all 43 planned conditions, including external tasks, using their archived category assignments. We display the archived ``Concept-to-factory'' family as ``Production handoff'' to describe the evaluated endpoint, without changing its membership. All Astra runs complete; Claude's eight timeouts occur on the first turn. Small category counts and differing tools preclude attributing the pattern to model capabilities alone.

\begin{table}[ht]
\centering\small
\caption{Timeouts / planned conditions by task family.}
\label{tab:category}
\begin{tabularx}{\linewidth}{@{}Yrr@{}}
\toprule Family & Astra & Claude\\\midrule
Brand identity & 0/6 & 1/6\\
Existing-brand campaign & 0/5 & 1/5\\
Packaging/SKU family & 0/3 & 0/3\\
E-commerce & 0/5 & 0/5\\
Multi-format adaptation & 0/4 & 2/4\\
Client-feedback revision & 0/5 & 1/5\\
Reference-to-original & 0/8 & 1/8\\
Production handoff & 0/7 & 2/7\\\midrule
Total & 0/43 & 8/43\\\bottomrule
\end{tabularx}
\end{table}

\section{Case-evidence provenance and reinterpretation}
The direct inspections read the actual last-turn responses in addition to archived judgments. For both systems in MFA-003 and CFR-002, these responses explicitly disclose an unresolved change and request a client decision. We therefore do not infer from these cases that the agent claimed the final delivery was complete. Runtime completion and an incomplete delivery remain observed; whether escalation was justified remains unadjudicated.

MFA-003 has no business visual in either system's \texttt{outputs/v3}; Claude does have a diagnostic image under \texttt{evidence}, which is excluded from the deliverable count. CFR-002's Astra preview has the same SHA-256 as its source. The Claude preview's changed-pixel bounding box is $[185,296,741,502)$, independently recomputed on the full RGB image; this localization does not measure exact title-scale accuracy or aesthetic acceptability.

The retained v0.2 judge-validation document records both rejected tool-judge controls and the artifact-only recheck. It explicitly states that the temporary tool-judge v1 result was discarded. The control figure reports that document's numbers and interventions, not reconstructed raw outputs. The case inspections perform no new counterfactual or judge run. The six additional inspections retain all original judgments, including the unresolved ECO-001 applicability discrepancy.

\clearpage
\section{Core conditions with archived semantic flags}
\label{app:flags}
All 18 flagged system-condition records in Table~\ref{tab:flags} have completed execution and full structural artifact-contract credit. Conditions sharing a brief are dependent. These flags preserve model judgments, including disputed or potentially justified blocked outcomes; they are not a count of false completion claims.

\begin{table}[!ht]
\centering\small\setlength{\tabcolsep}{4pt}
\caption{Any \texttt{VIOLATED} semantic check. Z/R/P denote context as in Appendix~\ref{app:brief}; DAB- is omitted.}
\label{tab:flags}
\begin{tabularx}{\linewidth}{@{}Yll@{}}
\toprule Core condition & Astra & Claude\\\midrule
L4-BID-001 (Z) & Flag & Flag\\
L4-BID-001 (R) & Flag & Timeout\\
L4-BID-001 (P) & Flag & No flag\\
L4-CAM-001 (Z) & Flag & No flag\\
L4-CAM-001 (P) & No flag & Flag\\
L4-PSF-001 (R) & Flag & No flag\\
L4-ECO-001 (R) & No flag & Flag\\
L4-MFA-002 (R) & Flag & Timeout\\
L4-MFA-003 (R) & Flag & Flag\\
L3-CFR-002 (R) & Flag & Flag\\
L4-RTO-001 (Z) & Flag & No flag\\
L4-RTO-001 (R) & Flag & No flag\\
L4-RTO-001 (P) & Flag & No flag\\
L4-CFRY-002 (R) & Flag & No flag\\
L4-CFRY-003 (R) & Flag & No flag\\\bottomrule
\end{tabularx}
\end{table}

\clearpage
\section{All 43 Brief Bank conditions}
\label{app:brief}
Table~\ref{tab:brief-paired} gives the common task inventory, including every Claude timeout. Z/R/P denote zero-shot, reference-grounded, and personalized contexts. Core intervals are missing-evidence bounds; external values are weighted scores. All Astra conditions completed. A Claude timeout is an execution outcome with no eligible quality score, not a zero score or a task awaiting execution.

\begin{table}[!ht]
\centering\footnotesize\renewcommand{\arraystretch}{1.0}
\caption{Evaluation v0.4 on Brief Bank data v0.3. DAB- prefixes are omitted. A completed execution does not guarantee a complete semantic delivery.}
\label{tab:brief-paired}
\begin{tabular}{@{}llrrl@{}}
\toprule Task & Context & Astra quality & Claude quality & Claude execution\\\midrule
\BriefPairedCaseRows
\bottomrule
\end{tabular}
\end{table}

\clearpage
\section{Additional visual evidence}
\label{app:visual-cases}
The six additional cases supplement the MFA-003 and CFR-002 inspections. Their exact prompts, final replies, inventories, original checks, direct file facts and 76 source hashes are retained in the project archive; 23 source PNGs are preserved byte-for-byte. The panels below complement the three main-text examples. These are selected illustrations, not an independent test set or estimates of failure frequency.

\begin{figure}[!ht]
\centering
\includegraphics[width=0.95\textwidth]{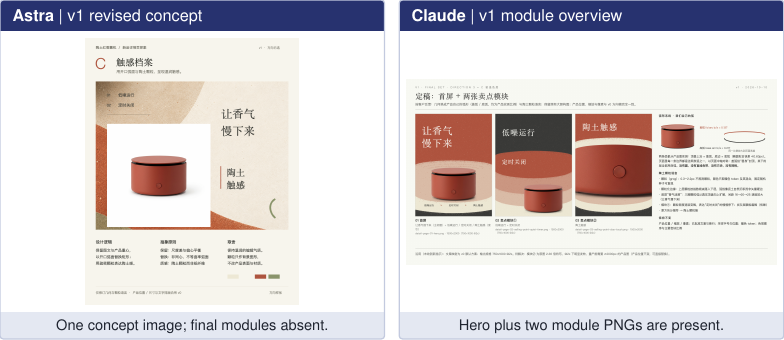}
\caption{RTO-001: Astra's revised concept and Claude's overview of the three requested modules. The inventory confirms three named module PNGs for Claude. Its response discloses uncertainty about the selected direction; this comparison does not adjudicate originality or that ambiguity.}
\label{fig:reference}
\end{figure}

\begin{figure}[!ht]
\centering
\includegraphics[width=0.95\textwidth]{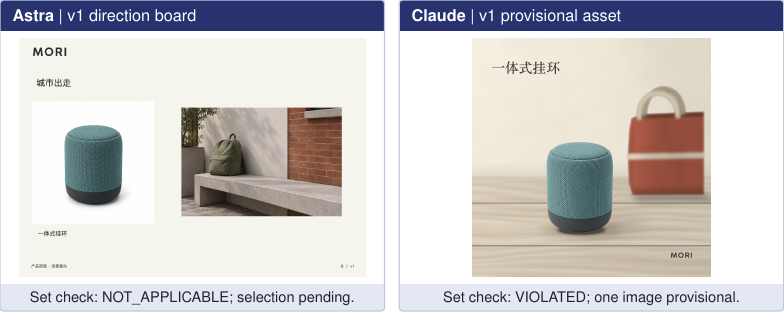}
\caption{ECO-001 evaluator-state discrepancy. Both candidates receive the same explicit direction selection. The left output is a revised concept; the right is one of six exported images, explicitly provisional. The associated complete-set judgments use different applicability decisions. This requires expert re-review and does not establish a revised system ranking; frozen scores are retained.}
\label{fig:applicability}
\end{figure}

\clearpage
\begin{figure}[!ht]
\centering
\includegraphics[width=0.95\textwidth]{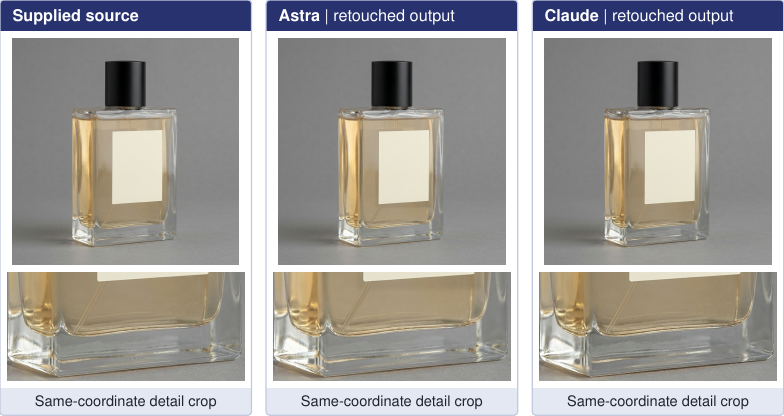}
\caption{ECO-002 positive contrast. Both systems use the same source and produce same-size PNGs; their archived checks contain no violation. Bottom panels show identical-coordinate crops, $x=270{:}750$, $y=730{:}950$, from the source and outputs. Model-judge bounds and visible changes do not replace expert judgments of preservation or retouch quality.}
\label{fig:retouch}
\end{figure}

\begin{figure}[!ht]
\centering
\includegraphics[width=0.95\textwidth]{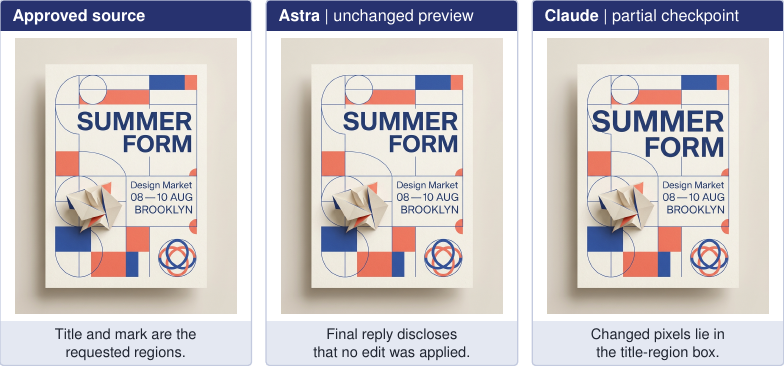}
\caption{CFR-002. The source and actual previews show a title enlargement request and an unresolved mark relocation. Astra's preview is byte-identical to the source; Claude's changed pixels lie within $[185,296,741,502)$ in the 928$\times$1152 image. Both final responses disclose a blocker. Pixel facts do not adjudicate whether stopping was justified or the partial edit was acceptable.}
\label{fig:cases}
\end{figure}

\end{document}